\documentclass[10pt,journal,final,twoside]{IEEEtran}
\usepackage{cite}
\ifCLASSINFOpdf
  \usepackage[pdftex]{graphicx}
  \graphicspath{{../pdf/}{../jpeg/}}
  \DeclareGraphicsExtensions{.pdf,.jpeg,.png}
\else
  \usepackage[dvips]{graphicx}
  \graphicspath{{../eps/}}
  \DeclareGraphicsExtensions{.eps}
\fi
\usepackage{amsmath}
\usepackage{amsfonts}
\usepackage[dvipsnames]{xcolor}
\usepackage{amssymb}
\usepackage{array}
\usepackage{multirow}
\usepackage{longtable}
\usepackage[english]{babel}
\usepackage{amsthm}
\usepackage{scalerel}
\usepackage{hyperref}

\usepackage{textcomp}
\usepackage{url}
\usepackage[
linesnumbered,ruled,vlined]{algorithm2e}
\usepackage {algpseudocode}
\usepackage{algorithmicx}
\usepackage{algcompatible}
\input{mysymbol.sty}
\makeatletter
\newcommand{\thickhline}{%
    \noalign {\ifnum 0=`}\fi \hrule height 1pt
    \futurelet \reserved@a \@xhline
}
\newcolumntype{"}{@{\hskip\tabcolsep\vrule width 1pt\hskip\tabcolsep}}

\makeatother
\ifCLASSOPTIONcompsoc
  \usepackage[caption=false,font=normalsize,labelfont=sf,textfont=sf]{subfig}
\else
  \usepackage[caption=false,font=footnotesize]{subfig}
\fi
\ifCLASSOPTIONcaptionsoff
  \usepackage[nomarkers]{endfloat}
 \let\MYoriglatexcaption\caption
 \renewcommand{\caption}[2][\relax]{\MYoriglatexcaption[#2]{#2}}
\fi
\let\MYorigsubfloat\subfloat
\renewcommand{\subfloat}[2][\relax]{\MYorigsubfloat[]{#2}}
\begin{document}
\title{Learning the Weather–Grid Nexus via \\ Weather-to-Voltage (W2V) Predictive Modeling}

\author{Sol Lim,~\IEEEmembership{Student~Member,~IEEE,} Min-Seung~Ko,~\IEEEmembership{Member,~IEEE,} Farnaz~Safdarian,~\IEEEmembership{Senior~Member,~IEEE,} and~Hao~Zhu,~\IEEEmembership{Senior~Member,~IEEE}}

\markboth{Submitted for Publications IEEE Transactions on Sustainable Energy}{}
\maketitle
 
\begin{abstract}
This paper proposes a weather-to-voltage (W2V) predictive modeling framework to learn the underlying weather–grid nexus. Unlike existing approaches on weather-informed grid operations, our proposed W2V model can achieve the joint analysis of weather and grid states, and further leverage this coupling to enhance grid-aware weather forecasting (GAWF) as a key application. 
To achieve this end-to-end learning, the W2V model acts as a differentiable surrogate for weather-incorporated power flow analysis by mapping weather features at high spatial resolution directly to grid-wide bus voltages. Thanks to a compact neural network design and principal component analysis based initialization, it achieves high  voltage prediction accuracy and numerical stability during training. Building on this capability, W2V-based voltage signals are used to guide the development of GAWF that can account for its downstream voltage prediction performance. Using a 6717-bus Texas synthetic test system with meteorological inputs from 701 weather locations, our numerical tests have verified the excellent accuracy and generalizability of the proposed W2V model. More importantly, the W2V model has enabled the GAWF to effectively prioritize the weather features and conditions that are most critical to grid operations, such as system-wide quick wind drops preceding ramp-ups.
\end{abstract}

\section{Introduction}
\label{sec:intro}
Weather conditions continue to grow as a crucial external factor in power system operations due to both the direct infrastructure impact and the indirect coupling with renewable energy sources (RES). Wind and solar generation capacity in the U.S.~tripled from 2014 to 2023, reaching more than 14\% of its total electricity generation \cite{EIA_EPA2023, EIA_EPA2014}, with similar upward trends worldwide. However, an increasing reliance on RES poses significant challenges on daily grid operations. Rapid fluctuations in weather conditions can cause voltage instabilities \cite{vittal2009steady}, while sharp increases in renewable output during favorable weather can quickly overload transmission lines  \cite{lannoye2014transmission}. Therefore, it is essential to improve the modeling and analysis of weather–grid nexus to ensure reliable and safe grid operations under increasing weather variability \cite{liu2018frequency, negnevitsky2014risk}.
%{\hao this paragraph should bring up the concept of weather-grid nexus, or weather-informed power system operations.}

To tackle these challenges, recent research and development efforts have advocated to incorporate weather information into grid modeling and operations. There is a growing interest in weather-based demand and renewable forecasts \cite{zhang2025weather, unlu2024weather}, as critical inputs for operational tasks such as day-ahead scheduling and real-time dispatch \cite{sweeney2020future}. Independent system operators (ISOs) such as MISO and ERCOT routinely rely on these forecasts to proactively manage operational uncertainty \cite{knueven2023stochastic,jascourt2023solar}. Meanwhile, AI advances are used to improve renewable forecast to mitigate weather-induced risks \cite{wang2016quantifying, yu2024short, zhang2025ultra}. While these efforts are well motivated by the dependence on weather conditions, they mainly focus on the forecast stage itself without accounting for the impact on downstream grid operational tasks. Unfortunately, this siloed framework fails to provide a full assessment of weather's impact on grid operations.   %As a result, standard forecasting metrics alone cannot fully assess the operational impact of weather on the grid.

% However, current approaches remain constrained by two fundamental limitations from a grid operational perspective. First, the weather information produced by conventional forecast models is essentially raw meteorological data. Hence, it provides little insight into the refined and operationally meaningful signals most relevant for grid operations. {\hao I dont fully understand the first issue - how does the earlier work above address it?} Second, the integration of this data is indirect and decoupled: by channeling weather information almost exclusively through RES forecasting, current approaches often fail to capture how weather directly influences critical grid states such as line flows and bus voltages. To further illustrate this second limitation, we present a motivating example in the following discussion. {\hao are we only saying current approaches decouple weather analytics with their grid impact. I don't think there is a need to have a very long discussion as the current one now. To me, it suffices to argue for the joint design because the impact of weather data cannot be evaluated using its own metric alone, which is what the example for.  }

\noindent {\bf Motivating Example:} To demonstrate this limitation, we present an example case study for which the current forecast-only framework is insufficient. Consider the simple two-bus system in Fig.~\ref{fig:motiv_diagram}, where the impact of renewable forecast error on transmission line congestion is analyzed using the DC power flow model. Depending on the weather condition and accordingly, the renewable generation level, the same 5-MW under-forecast error has a significant difference on line congestion. For the scenario of low RES output (10 MW), conventional generation is heavily relied on to meet the power demand, causing the transmission line to operate near its 30-MW capacity limit. This stressed grid operating condition makes a 5-MW RES forecast error to predict a 16.7\% overloading of the line. However, the same error has no impact at all on the line congestion for the scenario of high RES output (60 MW). This contrast underscores the limitations of a decoupled forecast framework, since forecast-only metrics fall short in capturing the impact on downstream grid operations.

\begingroup
\setlength{\textfloatsep}{4pt}
\begin{figure}[t!]
\vspace*{-8pt}
\centering
\includegraphics[width=0.95\columnwidth]{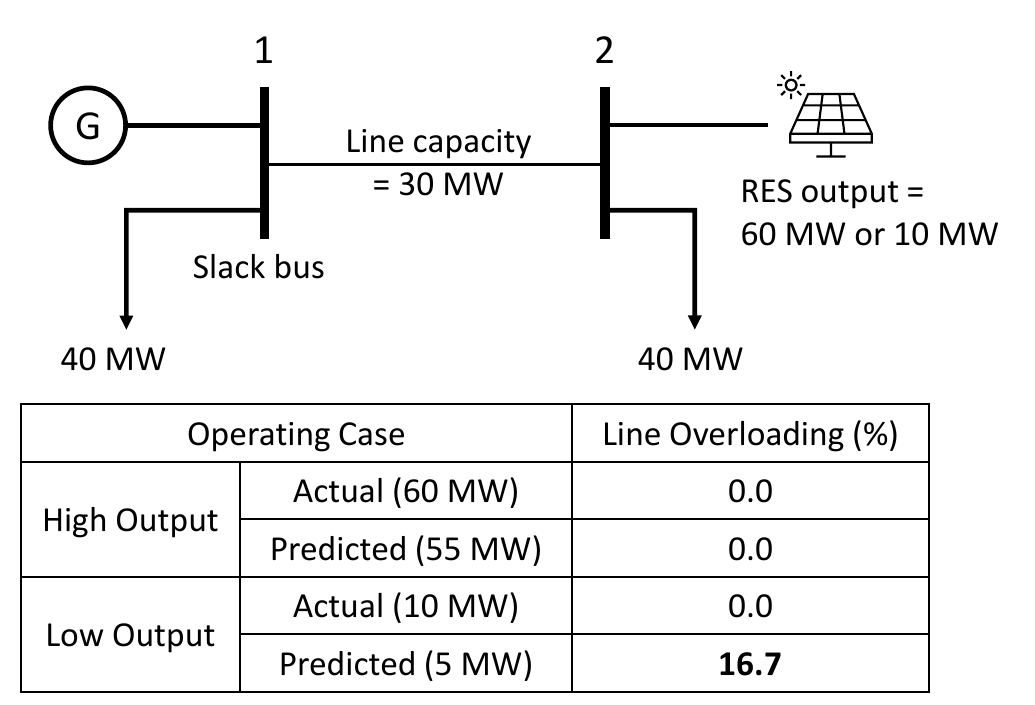}
\caption{Line overloading comparisons on a two-bus system.}
\vspace{-5pt}
\label{fig:motiv_diagram}
\end{figure}
\endgroup

% \begin{table}[!t]
% \renewcommand{\arraystretch}{1.3}
% \caption{Line Overloading Comparisons}
% \label{tab:motiv_results}
% \centering
% \setlength{\tabcolsep}{6pt}
% \begin{tabular}{|l|c|c|}
% \hline
% \multicolumn{2}{|c|}{Operating Case} & Line Overloading (\%) \\
% \hline
% \multirow{2}{*}{High Output} & Actual (60 MW) & 0.0 \\
% \cline{2-3}
%  & Predicted (55 MW) & 0.0 \\
% \hline
% \multirow{2}{*}{Low Output} & Actual (10 MW) & 0.0 \\
% \cline{2-3}
%  & Predicted (5 MW) & \textbf{16.7} \\
% \hline
% \end{tabular}
% \end{table}

%{\hao the nexus should be moved earlier and shortened. i think this is clearly driving what you mentioned in second paragraph. we need to directly present our own work earlier. }
%This observation underscores the limitation of current approaches that decouple weather from its grid impact, despite the inherent weather–grid nexus. Therefore, effectively embedding this coupling into modeling tasks would allow for grid awareness, aligning the objectives of models (e.g., weather forecasting) with the ultimate goal of reliable grid operation. 
Hence, it is crucial to embed the weather-grid coupling in the design of weather-dependent forecast approaches to enable their awareness of the grid operational impact. However, implementing such a grid-aware framework poses a key technical challenge: to train the forecaster with downstream grid objectives, a differentiable path from weather inputs to grid states is needed to guide the supervised learning process so that grid state gradients can back-propagate to the forecaster during training. This end-to-end learning process allows the forecaster not only to improve meteorological accuracy but also to internalize operationally meaningful signals for grid operation \cite{donti2017task, wahdany2023more, dvorkin2023price, beichter2024decision, 11122623}. 
%{\hao  better clarify the differentiable need.}
%{\hao  this part should belong to earlier paragraph. it's so brief and doesn't feel to need a full paragraph.}
Existing approaches to incorporate weather conditions into power flow analysis are based on individual scenarios and do not provide the aforementioned differentiable path. For example, component-level approaches for dynamic line rating \cite{ahmed2019weather, karimi2018dynamic} adjust line parameters for local weather conditions, yet are unable to capture any grid-wide effects. System-level approaches \cite{overbye2023approach, safdarian2024calculation, safdarian2025power} have successfully incorporated weather information into simulator-based power flow analysis using weather-dependent renewable models. Although this simulator-based approach can generate multiple scenarios from historical weather inputs, it remains necessary to develop a predictive model that represents these scenarios and can be seamlessly integrated into other supervised learning models. 
%fore, we need a differentiable alternative that preserves weather–grid fidelity while providing gradients from grid states back to weather inputs.

To address this gap, we put forth a new predictive modeling framework that can directly map from weather features to power system states. Specifically, we establish a weather-to-voltage (W2V) neural model trained on realistic or real-world datasets, providing an efficient, data-driven representation of weather-incorporated power flow scenarios. This surrogate model can be conveniently integrated with the forecast stage of weather data analytics, forming the core of our proposed weather-grid nexus learning. We demonstrate its applicability by developing a \textit{grid-aware weather forecaster (GAWF)}, where W2V-based bus voltages serve as a meaningful regularizing term during end-to-end training. Unlike existing siloed approaches that provide \textit{grid-unaware} weather forecast, our proposed GAWF directly informs weather modeling with its impact on grid states and thus offers critical information to downstream grid operations.

Our main contributions are summarized as follows:
\begin{enumerate}
    \item We develop a weather-to-voltage (W2V) model as a differentiable surrogate for simulator-based methods, that can be seamlessly integrated into end-to-end learning at the weather-grid nexus.

    \item We design a compact architecture for the W2V neural model and initialize it using principal component analysis, to improve the accuracy and numerical stability of the training process.
    %{\hao since numerical stability is a key advantage identified by experiments, was it emphasized in Sec II? - Updated.}   
    
    \item We develop a grid-aware weather forecasting framework that embeds W2V-based weather–grid coupling into the forecast stage. Specifically, grid state information is directly used as a regularizer to guide the weather forecaster in end-to-end learning from weather features to power flow outputs. 
    
    \item We demonstrate that our grid-aware weather forecaster (GAWF) offers operationally critical weather forecasts that account for downstream voltage levels, in comparison with the conventional, grid-unaware weather forecaster.
\end{enumerate}

The remainder of this paper is organized as follows. Section~\ref{sec:w2v} presents the design and training of our proposed W2V model. Section~\ref{sec:w2v_GAWF} discusses the use of W2V for developing GAWF. Section~\ref{sec:exp_setup} describes the experimental setup. Section~\ref{sec:numerical_results} provides a comprehensive analysis of the proposed W2V model and GAWF framework, and Section~\ref{sec:conclusion} concludes the paper.
%{\hao the introduction should be within 1.5 pages. Currently, the reasons for doing grid-informed weather analysis or W2V are both too long, and need to be more streamlined.}

\section{Weather-to-Voltage (W2V) Predictive Modeling}
\label{sec:w2v}
Our proposed W2V predictive model is designed to directly map spatial meteorological features to grid-wide bus voltage levels. This end-to-end prediction utilizes a compact autoencoder-based model and will enable the seamless integration with upstream/downstream tasks. 

To develop the W2V model, one uses a dataset denoted by $\mathcal{D} = \{ (\mathbf{w}_t,~\mathbf{v}_t)\}_{t=1}^T$, consisting of $T$ temporal samples. Each sample $t$ includes the following:
\begin{enumerate}
    \item Input weather features: vector $\mathbf{w}_t \in \mathbb{R}^W$ contains $N$ weather features such as temperature, wind speed, and global horizontal irradiance (GHI), measured at $S$ geographical locations, with the dimension $W = N \times S$.
    \item Output voltages: vector $\mathbf{v}_t \in \mathbb{R}^B$ represents the corresponding grid-wide bus voltages, with $B$ as the number of buses in the power system.
\end{enumerate}
These samples are time-synchronized and geographically related for the same time period and geo-regions. For weather features, the spatial resolution needs to be sufficiently high to ensure an accurate prediction. Our numerical studies later will use the \textbf{ERA5 reanalysis dataset} \cite{hersbach2023era5}, which provides high-resolution, geographically-specific meteorological data. However, this fine resolution inevitably leads to a high-dimensional $\mathbf{w}_t$ and increases the complexity of W2V modeling. Meanwhile, the proximity among the geo-locations implies that there is a strong correlation among weather features, which will be exploited for the W2V model design. 

To generate $\mathbf{v}_t$ from a given $\mathbf{w}_t$, power system simulators can be used. For example, weather conditions can be directly incorporated into renewable models to compute generation outputs, which are then used to produce steady-state power flow solutions \cite{overbye2023approach, safdarian2024calculation}. Alternatively, grid operators can generate $\mathcal{D}$ by integrating historical weather data with time-synchronized grid operational data. Note that we assume that the output bus voltages represent the quasi-steady-state response of power systems, with no dynamic or transient effects. Hence, the samples are not temporally related, and each sample represents a distinct operating point driven by evolving weather conditions. The detailed description of the dataset used for our numerical validations is presented in Section~\ref{sec:exp_setup}. 

Our W2V predictor aims to obtain a mapping, $\pi_v$, to match the weather–grid interactions, given by:
\begin{align}
    \hat{\mathbf{v}}_t = \pi_{v}(\mathbf{w}_t;\boldsymbol{\theta})
    \label{eq:w2vmap}
\end{align}
where $\boldsymbol{\theta}$ denotes the NN model parameters. Instead of using a generic NN model, we take advantage of the inherent spatial correlation within the weather data to design a compact autoencoder (AE) model. The gist of AE is to learn a compressed representation of input data by utilizing two sub-networks: an encoder $E$ and a decoder $D_w$. The encoder maps the high-dimensional input vector $\mathbf{w}_t$ to a low-dimensional latent space with vector $\mathbf{z}_t := E(\mathbf{w}_t) \in \mathbb{R}^{K}$. The decoder then attempts to reconstruct the original input, $\hat{\mathbf{w}}_t$, from the latent $\mathbf{z}_t$. Using a simple single-layer AE, we can express this process as:
\begin{align}
    \mathbf{z}_t &= E(\mathbf{w}_t) = \sigma(\mathbf{W}_{\text{enc}}\mathbf{w}_t + \mathbf{b}_{\text{enc}}) \\
    \hat{\mathbf{w}}_t &= D_w(\mathbf{z}_t) = \sigma(\mathbf{W}_{\text{dec}}\mathbf{z}_t + \mathbf{b}_{\text{dec}})
\end{align}
where $\mathbf{W}$ and $\mathbf{b}$ are the weight and bias parameters included by $\boldsymbol{\theta}$, and $\sigma$ is the activation function. The AE model can be trained by minimizing the reconstruction error, typically the mean squared error (MSE) given by $\mathcal{L}_\text{w} = \| \mathbf{w}_t - \hat{\mathbf{w}}_t \|^2$. In short, the strength of AE lies in the introduction of a low-dimensional latent space to effectively extract compact yet expressive features from high-dimensional input \cite{hinton2006reducing}. This powerful dimensionality reduction technique has proven effective in many domains dealing with spatially correlated data, such as video compression \cite{habibian2019video}, medical imaging analysis \cite{galic2023machine}, and weather data analysis \cite{kurihana2024identifying}.

\begin{figure}[t]
    \centering    
    \includegraphics[width=0.45\textwidth]{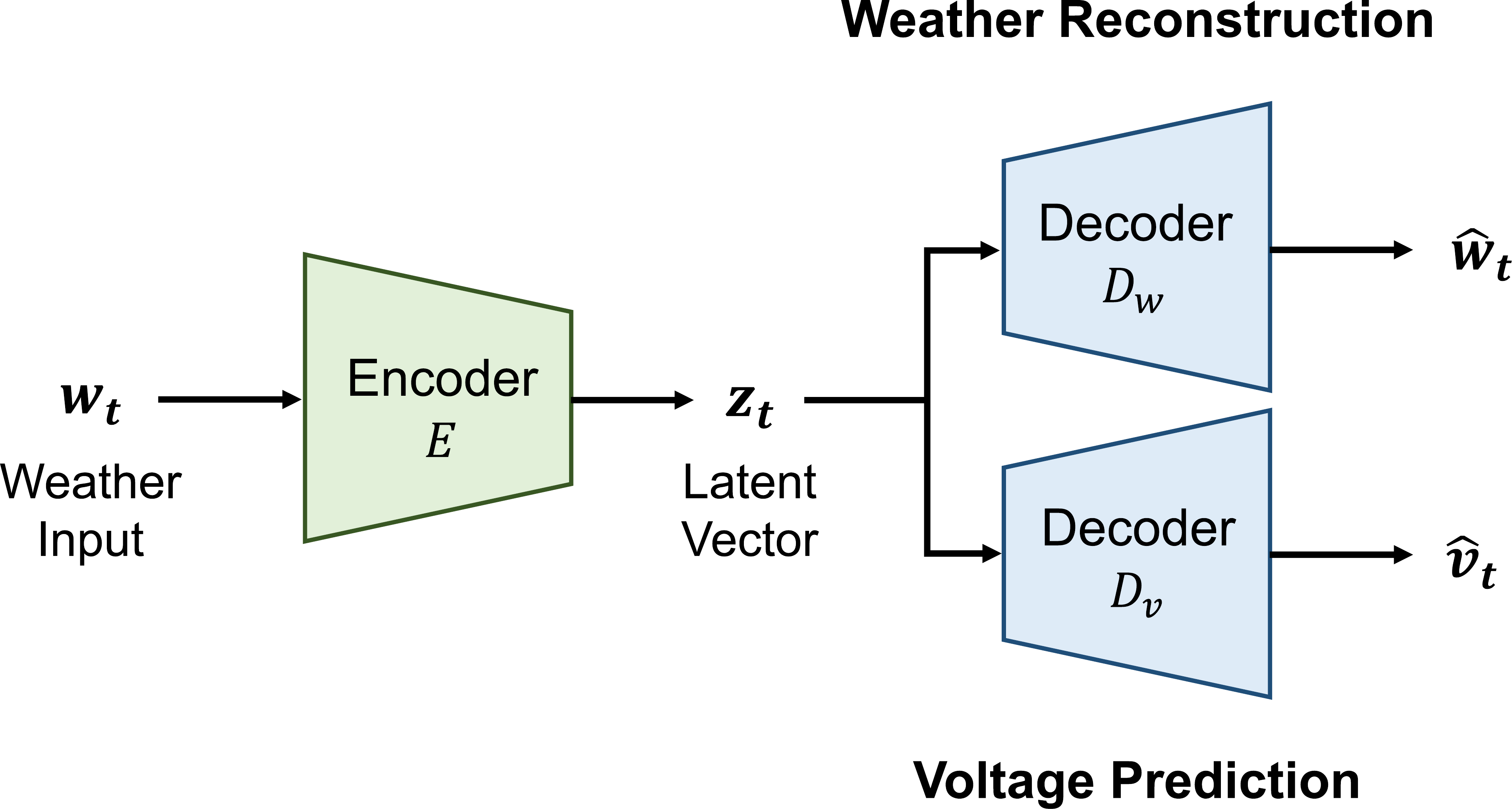}
    \caption{Autoencoder-based W2V model architecture. An encoder $E$ compresses the weather input $\mathbf{w}_t$ to a latent vector $\mathbf{z}_t$, which then drives two parallel decoders. The primary decoder $D_v$ predicts the grid-wide bus voltages in $\hat{\mathbf{v}}_t$, while the secondary decoder $D_w$ reconstructs the weather input $\hat{\mathbf{w}}_t$. The W2V parameters $\boldsymbol{\theta}$ include all weights/biases in the encoder and two decoders, and are jointly trained using the dataset $\mathcal{D}$.}
    \label{fig:mapper}
\end{figure}

Building upon the AE principle, we design a special W2V architecture consisting of a single encoder and dual decoders, as illustrated in Fig.~\ref{fig:mapper}. In addition to the standard encoder $E$ and decoder $D_w$ to reconstruct $\mathbf{w}_t$, our W2V includes another decoder, $D_v$, to perform the main predictive task given by $\hat{\mathbf{v}}_t := D_v(\mathbf{z}_t)$. 
Note that the original decoder $D_w$ can assist the W2V prediction as a regularizer by improving the compactness and robustness of weather compression by the encoder. This, in turn, prevents the latent vector from overfitting to the voltage prediction task and enhances the overall generalizability of our W2V model.

To train the dual decoders, we minimize the following regularized loss function: 
\begin{equation}
    \mathcal{L}
   = \mathcal{L}_\text{v}(\hat{\mathbf{v}}_{t}, \mathbf{v}_{t})+ \lambda\mathcal{L}_\text{w}(\hat{\mathbf{w}}_{t}, \mathbf{w}_{t}),
\end{equation}
where $\mathcal{L}_\text{v}$ denotes the MSE for predicting the voltage profiles. It is regularized by the MSE for weather reconstruction, $\mathcal{L}_\text{w}$, using the hyperparameter $\lambda > 0$. A smaller $\lambda$ prioritizes the voltage prediction accuracy, while a larger value emphasizes the preservation of weather information. Given this critical role, we determine the best value for $\lambda$ through trade-off analysis. Specifically, we train the W2V model across a range of $\lambda$ values to analyze the trade-off between the two MSE terms. A good choice of $\lambda$ should attain a substantial reduction in $\mathcal{L}_\text{w}$ without significantly compromising the core task of voltage prediction. This combined training approach, a form of multi-task learning that leverages an auxiliary task for regularization \cite{ruder2017overview}, ensures the learned latent representation remains both voltage-aware and meteorologically informative to improve the overall model generalizability.

To ``warm-start'' the encoder, we use the principal component analysis (PCA)-based projection as the initialization. The routine ``cold-start'' via randomized NN parameters is prone not only to computational inefficiency but also to unstable optimization. We address this issue by using the powerful PCA-based linear projection \cite{mishkin2015all}. Specifically, for each weather feature, our experimental analysis has shown that just a couple of principal components are sufficient to maintain the majority of weather information (at least $92\%$ of variance). This implies a substantial level of linear correlation and allows one to design the latent space dimension, $K$, to be in the order of a few variables. Thus, the encoder $E$ is initialized by the PCA-based transformation:
\begin{equation}
\mathbf{z}_t \leftarrow \mathbf{P}^\top \mathbf{w}_t- (\mathbf{P}^\top \boldsymbol{\mu}),
\label{eq:pca-init}
\end{equation}
with $\mathbf{P} \in \mathbb{R}^{W\times K}$ and $\boldsymbol{\mu}\in \mathbb{R}^{W}$ representing the projection matrix and the mean vector obtained by PCA, respectively. This initialization provides a simple, yet effective, linear baseline that also improves numerical stability during training. Further refinement of encoder parameters through AE training can identify more complex, nonlinear relations while adjusting to the voltage mapping, as shown by our experimental results. 

\section{W2V for Grid-Aware Weather Forecasting}
\label{sec:w2v_GAWF}

We present the use case of our W2V predictor in an upstream weather forecasting (WF) task by developing a grid-aware weather forecaster (GAWF). The idea is to pursue weather forecast that can produce more reliable and critical mapping to grid voltage states, instead of solely optimizing the forecast accuracy. %To clearly evaluate its effectiveness, we also introduce a conventional grid-unaware weather forecaster (GUWF) for comparison.
The latter is the conventional framework and will be referred to grid-unaware weather forecaster (GUWF) for comparison.

For every time instant $t$, both WFs seek to predict future weather data, $\mathbf{y}_t$, from the past weather input sequence, $\mathbf{x}_t$. For fairness, the same predictive model $\mathcal{F}(\cdot)$  is adopted:
\begin{align}
    \text{GUWF}:&\;\hat{\mathbf{y}}_{t} = \mathcal{F}(\mathbf{x}_{t};\bbphi) \\
    \text{GAWF}:&\;\hat{\mathbf{y}}'_{t} = \mathcal{F}(\mathbf{x}_{t};\bbphi'),
\end{align}
with the respective parameters $\bbphi$ and $\bbphi'$. We will implement $\mathcal{F}(\cdot)$ using the Transformer model, as detailed later. For a given forecast horizon $h$, one can determine the length of input sequence $k$ and the lead time $\ell$ by using a simple grid search based on the prediction performance, computational efficiency, and grid operational requirements \cite{ko2022feedforward}. Upon selecting $k$ and $\ell$, the input sequence and target output per time $t$ become $\mathbf{x}_t=(\mathbf{w}_{t-k+1},\ldots,\mathbf{w}_t)$ and $\mathbf{y}_{t}=(\mathbf{w}_{t+\ell},\ldots,\mathbf{w}_{t+\ell+h-1})$. Recall that the weather data sample $\mathbf{w}_t\in\mathbb{R}^{S\times{N}}$ was defined in Section~\ref{sec:w2v}. The forecasted weather outputs of GAWF and GUWF can be used to obtain the corresponding grid voltages based on the W2V model in \eqref{eq:w2vmap}. This way, the W2V predictor is assumed to have already been trained and thus its model parameters are fixed when analyzing the impact on the WF task.

Different from the conventional GUWF, the proposed GAWF advocates to introduce the voltage prediction to the overall training loss objective. First, the GUWF uses the standard  WF accuracy for its loss function,  
%which can be regarded as a common practice, the training loss only considers the weather forecasting error as:
%\begin{equation}
    $\mathcal{L}_{GU} = \mathcal{L}_\text{w}(\hat{\mathbf{y}}_{t}, \mathbf{y}_{t})$,
%\end{equation}
which is typically  defined by the norm of the prediction error $(\hat{\mathbf{y}}_{t}-\mathbf{y}_{t})$, such as the MSE.
%For the loss function $\mathcal{L}_\text{w}$, we adopt the MSE. 
Therefore, under GUWF weather forecasting and voltage mapping can be regarded separate tasks. Instead, we can directly incorporate the voltage prediction into the design of weather forecasters as an additional training loss , leading to the following loss for GAWF:
\begin{align}
    \mathcal{L}_{GA} = \mathcal{L}_\text{w}(\hat{\mathbf{y}}'_{t}, \mathbf{y}_{t}) + \gamma \mathcal{L}_\text{v}\left(\pi_v(\hat{\mathbf{y}}'_{t}), \mathbf{s}_{t}\right), \label{eq:galoss}
\end{align}
where the second term stands for the \textit{voltage loss} that aims to match the W2V-predicted voltages with the target voltages in $\mathbf{s}_{t} = (\mathbf{v}_{t+\ell},\ldots,\mathbf{v}_{t+\ell+h-1})$. 
%represents the target voltage sequence, $\hat{\mathbf{z}}'_{t}=\pi_v(\hat{\mathbf{y}}'_{t};\boldsymbol{\theta})$ corresponds to the predicted voltage sequence, and 
Here, the hyperparameter $\gamma$ will be chosen to balance the two loss terms, as detailed soon. 

\begin{figure}[t]
    \centering
    \begin{tabular}{@{}c@{}}
         \includegraphics[width=0.95\linewidth]{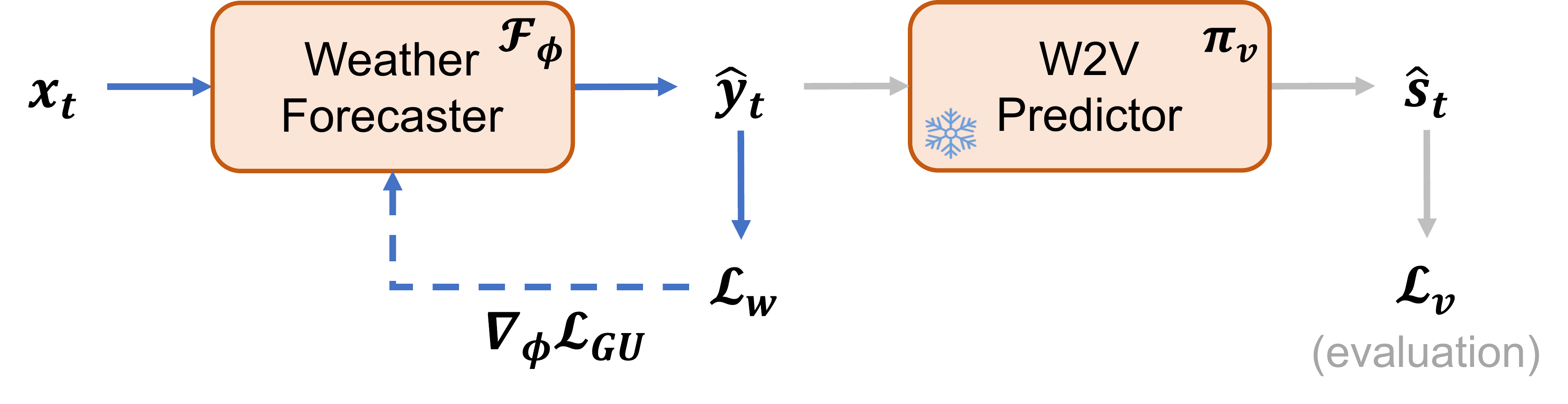} \\
         \small (a) \\
    \end{tabular}
    \begin{tabular}{@{}c@{}}
         \includegraphics[width=0.92\linewidth]{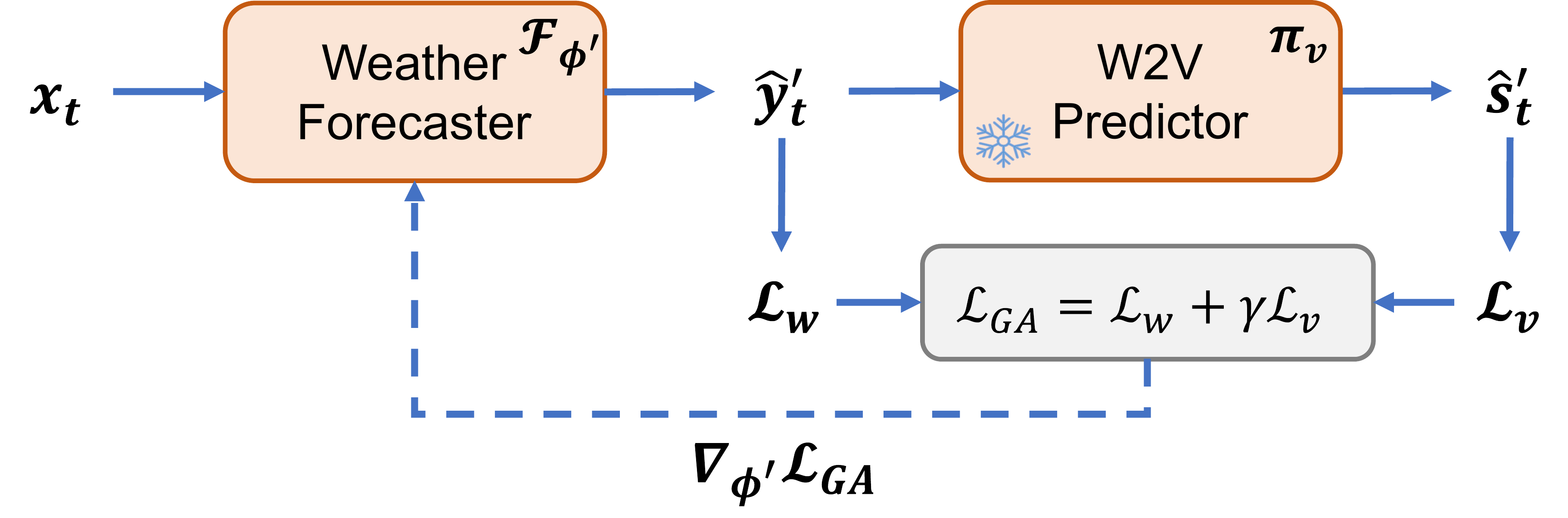} \\
         \small (b) \\
    \end{tabular}
    \caption{Training schematic of (a) GUWF and (b) GAWF.}
    \label{fig:training_flow}
    %\vspace{-0.2cm}
\end{figure}

Fig.~\ref{fig:training_flow} illustrates the overall architecture of the two models, of which GAWF jointly incorporates both loss functions. Thanks to the additional information provided by the voltage loss, the GAWF  is expected to benefit grid operations with more accurate voltage states. Of course, the improvement in voltage prediction could slightly sacrifice the WF performance. This trade-off can be controlled by setting an appropriate $\gamma$. As demonstrated by our numerical results later, we can select the best $\gamma$ that enhances the voltage prediction   at the price of a slight increase in the WF error. 
%compromising weather forecasting accuracy.

Meanwhile, training the GAWF has the same computational complexity as GUWF. Note that the trained GUWF model can serve as a good initialization to ``warm-start'' GAWF. This inspires us to develop  a \textit{two-stage} training procedure for GAWF. To establish a strong baseline, the first stage pre-trains the model as a GUWF using only the WF loss $\mathcal{L}_{\text{w}}$ for at least two-thirds of the GUWF total epochs to attain sufficiently high WF accuracy. %{\hao (maybe give an idea on number of epochs?)} 
Afterwards, the second stage uses the joint loss $\mathcal{L}_{GA}$ in \eqref{eq:galoss} to effectively reduce the voltage error until convergence. Note that the gradient update for $\mathcal{L}_{GA}$ can be efficiently computed using standard packages, because the W2V model $\pi_v$ is also a neural network that maintains the fast back-propagation capability. Hence, this approach can greatly accelerate the training of GAWF.
%the GAWF is initialized using parameters from a pre-trained GUWF, optimized solely on the weather forecasting loss $\mathcal{L}_{GU}$. This stage ensures the model first achieves high weather forecasting accuracy. In the second stage, we fine-tune the model by introducing a voltage prediction regularization term, employing a composite loss function $\mathcal{L}_{GA}$. This approach enables the GAWF to refine its parameters and produce forecasts better aligned with grid operational constraints.

As for the unified WF model, $\mathcal{F}(\cdot)$, we adopt a multi-modal Transformer architecture with the channel-independent design in \cite{Yuqietal-2023-PatchTST, li2024multi}. Unlike conventional recurrent models that process data sequentially, Transformers use a self-attention mechanism to simultaneously capture long-range dependencies and global temporal interactions, and this can effectively mitigate issues like vanishing gradients \cite{NIPS2017_3f5ee243}. The channel-independent design is particularly well-suited for our W2V predictor. Since each weather feature has a distinct effect on the downstream voltage prediction task based on weather-informed grid modeling in \cite{overbye2023approach,safdarian2024calculation}, it is useful to represent their unique temporal patterns. This way, our Transformer model employs modality-specific branches, each with its own self-attention mechanism, to effectively learn the distinct dynamics of each weather feature. Specifically, each branch consists of a vanilla Transformer encoder and a linear decoder. The input tensor $\mathbf{x}_{t} \in \mathbb{R}^{k \times S \times N}$ is split into $N$ matrices, one for each weather feature, for independent processing by the branch. The outputs also need to be concatenated to form the tensor $\hat{\mathbf{y}}_{t} \in \mathbb{R}^{h \times S \times N}$. To improve efficiency, we also use linear projection and learnable temporal embeddings as inputs to each branch. Furthermore, the encoder depth is chosen based on the forecast horizon $h$. Fewer layers are used for short-term and deeper encoders for long-term horizons, to better capture varying temporal dependencies.
Finally, it is important to note that our proposed grid-aware forecasting framework is model-agnostic, as the proposed methodology is flexible with a different model of $\mathcal{F}(\cdot)$. 
%is readily replaced with any other advanced architecture without altering the core methodology.

\section{Experimental Setup}
\label{sec:exp_setup}

%\textcolor{blue}{Overall, I feel like each paragraph is too short.}
This section presents the experimental setup used in numerical studies. We first introduce the weather-grid dataset and implementation details and then summarize   the hyperparameter settings for both the W2V predictor and the GAWF model.

\subsection{Dataset and implementation details}
\label{subsec:data}
\begin{figure}[tp]
  \centering
  \includegraphics[width=0.45\textwidth]{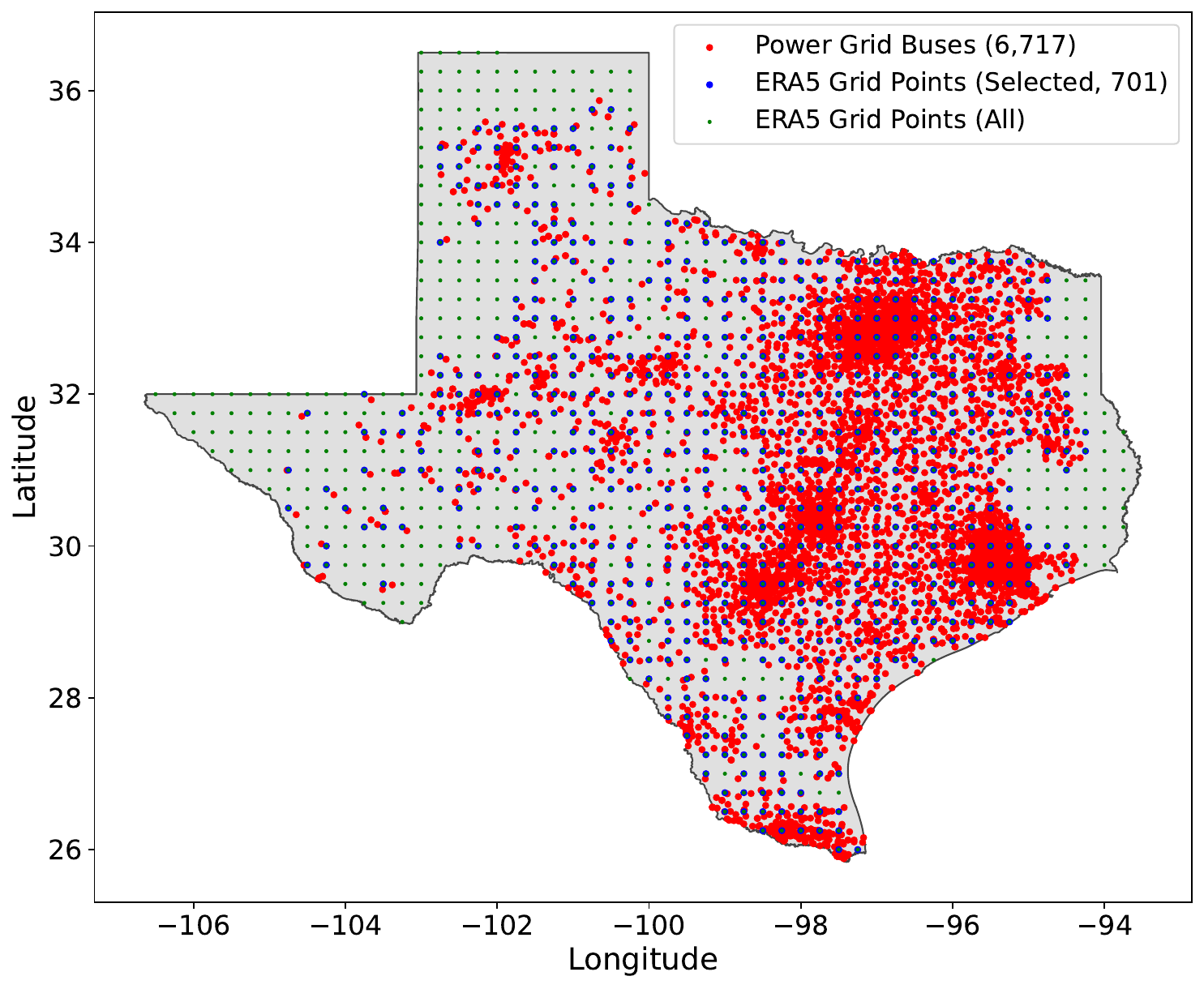}
  \caption{Spatial comparison of the selected 701 ERA5 grid points and the 6717 bus locations within the Texas synthetic grid.}
  \label{fig:WeatherData}
\end{figure}

The meteorological inputs are obtained from the ERA5 reanalysis dataset provided by the European Centre for Medium-Range Weather Forecasts (ECMWF) \cite{hersbach2023era5}. This is a widely-adopted weather dataset due to its fine spatial resolution of approximately 25--30~km. This level of resolution is suitable for capturing geographically diverse weather patterns over a large interconnection. For the Texas grid, we first identify 1046 ERA5 grid points within the grid boundary and then select $S=701$ grid points that are geographically closest to each power grid bus. This selection provides a high-resolution mapping between weather conditions and grid elements for downstream power system simulations. Fig.~\ref{fig:WeatherData} illustrates the geographical distribution of all ERA5 grid points, the selected 701 grid points closest to grid buses, and the bus locations in the Texas grid.

The voltage outputs used for the numerical validations are generated using the PowerWorld power flow solver on the synthetic 6717-bus Texas system. The hourly weather data for 2016 are mapped to the grid model to generate grid voltages following established procedures \cite{overbye2023approach, safdarian2024calculation}, resulting in a total of $T=8784$ samples. For each time step $t$, input vector $\mathbf{w}_t$ in \eqref{eq:w2vmap} consists of $N=3$ key meteorological features: temperature, wind speed, and global horizontal irradiance (GHI). Temperature is a primary driver for load behavior, while wind speed and GHI directly influence renewable generation. Wind speed is calculated by combining two directional components in ERA5, following the method in \cite{safdarian2024calculation}. Furthermore, we choose GHI instead of direct irradiance because the photovoltaic generation is more strongly correlated with the total solar energy incident on a surface. Accordingly, the weather input vector $\mathbf{w}_t$ has dimension $W = N \times S = 3 \times 701$. For each sample, the output voltage vector $\mathbf{v}_t$ has dimension $B=6717$ in per unit (p.u.). 

The weather-voltage dataset is filtered to remove unreliable samples. Specifically, 46 time steps with voltage magnitudes exceeding 1.20 p.u. are excluded, as these cases may experience potential solver convergence issues. The remaining samples are then normalized to the interval $[0, 1]$ using the min-max scaling . The normalized dataset is partitioned differently depending on the task. For the W2V predictor, the dataset is randomly divided into training (70\%), validation (20\%), and test (10\%) sets to evaluate generalization across various weather conditions without temporal bias. For the weather forecasting models (GUWF and GAWF), the same split ratios are applied chronologically to preserve the temporal structure inherent in the forecasting task.
All NN models are implemented using PyTorch \cite{NEURIPS2019_bdbca288} and trained using the RAdam optimizer \cite{Liu2020On} for up to 500 epochs with early stopping based on validation loss to prevent overfitting. The initial learning rate and batch size are selected via grid search, and a dynamic learning rate scheduler reduces the learning rate if the validation loss plateaus. The scheduler parameters are also tuned for each forecasting horizon to ensure optimal performance. For fair comparison, the same optimized hyperparameters are used for both grid-unaware (GU) and grid-aware (GA) models. All experiments are conducted on Google Colab environment.

\subsection{Hyperparameter settings}
\label{sec:hyper}
This subsection summarize the key hyperparameters for both the W2V predictor and WF models. Additional detailed trade-off analyses for tuning regularization hyperparameters are also presented.

\paragraph{W2V predictor} 
As discussed earlier, the AE model for W2V is configured with input/output dimensions of $W=2103$ and $B=6717$, respectively. The dimension $K$ of latent vector $\mathbf{z}_t$ is determined using PCA-based initialization. Specifically, $\{5, 21, 5\}$ principal components are adopted for temperature, wind speed, and GHI, respectively, such that the reconstruction error of each feature remains below 8\%. This results in a total latent dimension of $K = 5+21+5 = 31$. 

\begin{figure}[t]
    \centering
    \begin{tabular}{@{}c@{}}
         \includegraphics[width=\linewidth]{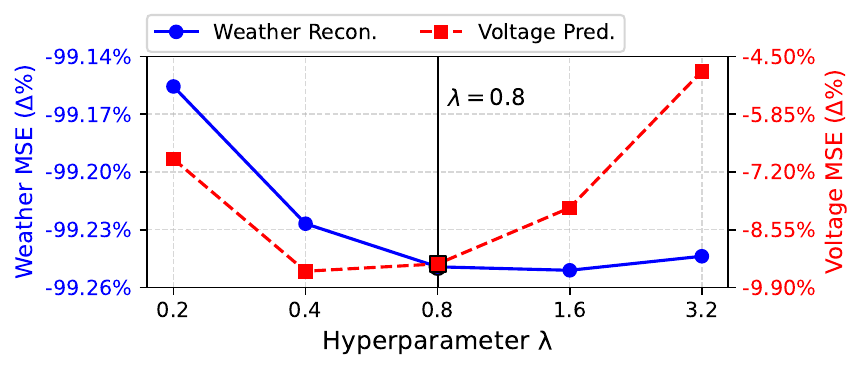} \\
         \small (a) W2V predictor ($\lambda$) \\
    \end{tabular}
    \begin{tabular}{@{}c@{}}
         \includegraphics[width=\linewidth]{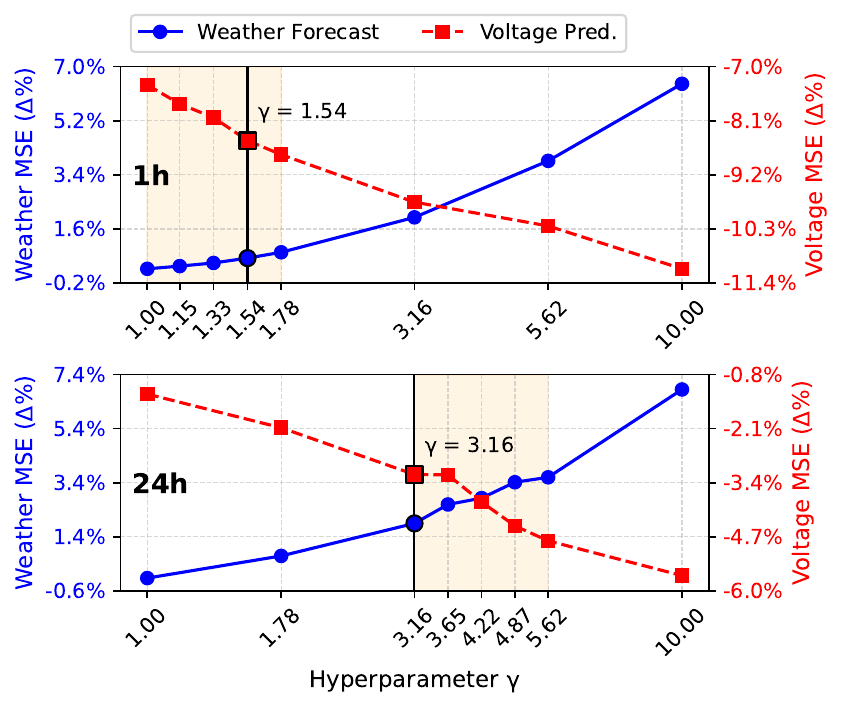} \\
         \small (b) GA forecaster ($\gamma$) \\
    \end{tabular}
    \caption{Trade-off analyses for selecting regularization hyperparameters: (a) W2V predictor ($\lambda$); (b) GA forecaster ($\gamma$). The vertical marker denotes the selected value; the shaded region in (b) indicates the refined search range.}
    \label{fig:tradeoff_analysis}
\end{figure}

To select the regularization hyperparameter $\lambda$, we evaluate geometrically-spaced candidates $\lambda \in \{0.2, 0.4, 0.8, 1.6, 3.2\}$. For each candidate, the percentage change in voltage prediction MSE ($\mathcal{L}_\text{v}$) and weather reconstruction MSE ($\mathcal{L}_\text{w}$) are evaluated relative to the baseline case with $\lambda = 0$. As shown in Fig.~\ref{fig:tradeoff_analysis}(a), $\lambda = 0.8$ achieves the best trade-off, providing sufficient regularization while maintaining strong voltage prediction performance.

\paragraph{WF models}
The MultiModalTransformer architecture is adopted for all WF experiments, as described in Section~\ref{sec:w2v_GAWF}. Its core hyperparameters, which remain constant throughout all experiments, are summarized in Table~\ref{tab:forecaster_params}. Our experimental evaluation covers five distinct forecasting horizons from 1 to 24 hours. To account for the increased complexity for longer-term forecasting, the number of Transformer layers ($n_{\text{layers}}$) is larger for 12-/24-hr horizons than for up to 6-hr. 

The GAWF regularization hyperparameter $\gamma$ is tuned separately for each forecasting horizon $h$ using a two-stage logarithmic search procedure. A coarse scan is performed over the interval $[1.0, 10.0]$, followed by a refined search around the best-performing region. The selected $\gamma$ values are $\{1.54, 2.37, 2.37, 3.16, 3.16\}$ for the $h=\{1, 3, 6, 12, 24\}$-hr horizons, respectively. Fig.~\ref{fig:tradeoff_analysis}(b) shows the trade-off analysis results for the 1-hr and 24-hr horizons. Averaged across all five horizons, the GAWF model reduces voltage prediction error by 4.562\% with only a minor 1.235\% increase in total weather forecast error, representing a highly favorable trade-off for voltage prediction accuracy.
%\textcolor{blue}{Have we mentioned that voltage prediction is the primary task?}

\begin{table}[t]
\centering
\caption{Hyperparameters for the Weather Forecast Model}
\label{tab:forecaster_params}
\renewcommand{\arraystretch}{1.4}
\begin{tabular}{lc}
\hline
Hyperparameter & Value \\
\hline
Model Dimension ($d_{\text{model}}$) & 128 \\
Number of Attention Heads ($n_{\text{heads}}$) & 8 \\
Number of Encoder Layers ($n_{\text{layers}}$) & 2 or 4 \\
Dropout Rate & 0.2 \\
Sequence Length & 48\\
\hline
\end{tabular}
\end{table}

\section{Numerical Results}
\label{sec:numerical_results}
%{\ms{May need to change the overall paragraph. Not aligned with the overall Section contents.}}
This section presents the numerical results of the proposed W2V model and examines its applicability to weather forecasting via the GAWF. We first validate the W2V prediction accuracy, and then assess the GAWF performance across different forecasting horizons and weather conditions, with a deeper analysis of large voltage error events. The results demonstrate that the W2V-enabled GAWF can use the predicted grid state information to better align weather forecasts with grid operation needs.

\subsection{W2V prediction performance}
\begin{figure}[t]
    \centering    
    \includegraphics[width=0.85\columnwidth]{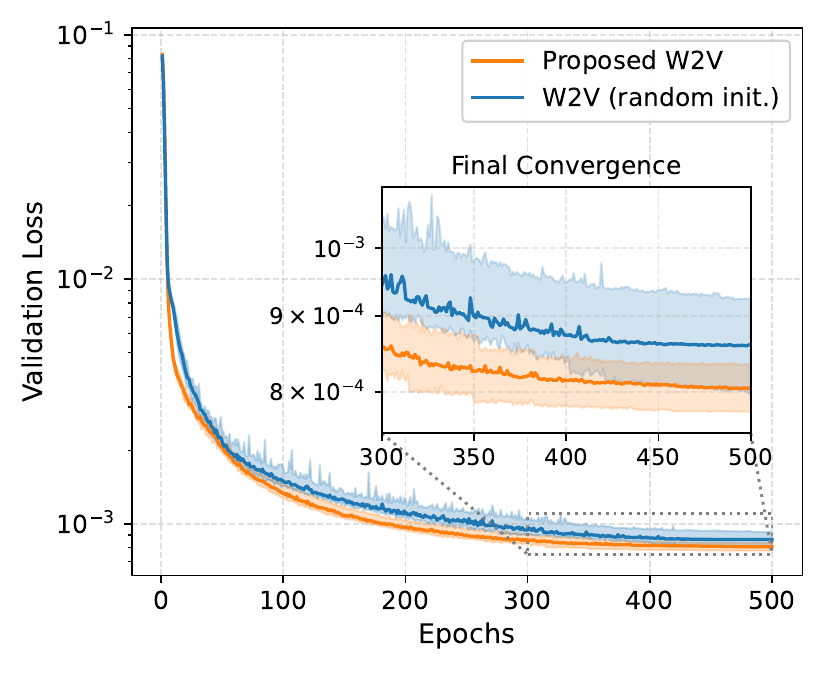}
    \vspace{-5pt}
    \caption{W2V validation loss comparison between random and PCA-based initialization.}
    %{\hao every setence needs a period; didn't explain the shaded region}
    \label{fig:initialization_comparison}
\end{figure}

%{\hao this part seems pretty straightforward. Could we provide some comparative study on W2V model developments?}
We present the W2V prediction performance on the test set defined in Section~\ref{subsec:data} 
%{\hao where was training/test divide discussed?-Revised in Section IV-A. We now explicitly state that W2V uses a random split, whereas GUWF/GAWF use a chronological split.} 
using the root mean-squared error (RMSE) and mean absolute error (MAE), both in p.u.~values. Using the hyperparameter $\lambda$ selected from Section~\ref{sec:hyper}, we compare the W2V model with two baseline approaches: i) a multilayer perceptron (MLP) with three hidden layers and a comparable number of parameters to the proposed W2V model,
%{\hao was \# of para mentioned earlier?}
and ii) an autoencoder with random initialization. These baselines allow us to quantify the benefits of the proposed NN design and PCA-based encoder initialization.    
%{\hao unclear} 
%We also quantify the benefit of PCA-based initialization in the proposed W2V by comparing it against a W2V model with a randomly initialized encoder. 
Table~\ref{tab:W2V_performance} lists the mean and standard deviation of the RMSE and MAE values in 10 random seeds. Even with random initialization, the autoencoder already outperforms MLP, achieving roughly 20\% error reduction in both metrics. Incorporating PCA-based initialization further improves the results, with a noticeable reduction in error variability across seeds. This robustness is also confirmed by Fig.~\ref{fig:initialization_comparison}, which compares the validation loss trajectories with the mean values in solid lines and the range over 10 seeds indicated by the shaded regions. Clearly, the proposed W2V's trajectory exhibits more stable convergence with smaller seed-to-seed variability.  
Together, these results demonstrate that the proposed W2V model design not only improves the prediction accuracy but also helps the numerical stability of the training process. 

\begin{table}[t]
\centering
\caption{W2V Voltage Prediction Performance Comparison}
\label{tab:W2V_performance}
\renewcommand{\arraystretch}{1.3}
\begin{tabular}{c|c|c}
\hline
Method & RMSE [$\times 10^{-3}$ p.u.] & MAE [$\times 10^{-3}$ p.u.] \\ \hline
MLP & $3.943 \pm 0.117$ & $2.148 \pm 0.052$ \\ 
\hline
AE (random init.) & $3.175 \pm 0.052$ & $1.691 \pm 0.030$ \\
\hline
Proposed W2V & $\textbf{3.112} \pm \textbf{0.043}$ & $\textbf{1.649} \pm \textbf{0.027}$ \\
\hline
\end{tabular}
\end{table}

\begin{figure}[t]
    \centering        
    \includegraphics[width=\columnwidth]{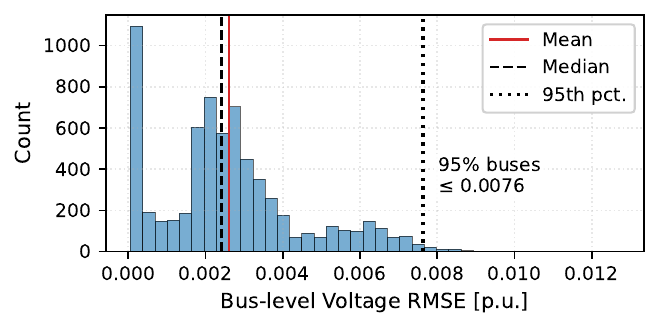}
    \vspace{-0.7cm}
    \caption{Histogram of bus-level voltage prediction RMSE across 6717 buses.} %\textcolor{red}{need figure update on high RMSE values?}
    \label{fig:bus-wise_error}
\end{figure}

To assess the consistency of the W2V prediction throughout the grid, Fig.~\ref{fig:bus-wise_error} presents the histogram of the bus-wise voltage RMSE for all 6717 buses. 
%{\hao you could have rephrased this bus-wise term to make it more concrete} 
%{\hao 95th not mentioned, do we need it?}
Most buses exhibit small RMSE values below 0.004~p.u., with the largest concentration near 0~p.u. The 95th percentile is at 0.0076~p.u., indicating that relatively large prediction errors are confined to a small subset of buses. Moreover, the mean and median RMSE values are very close to each other, indicating a balanced distribution of prediction errors across buses. Given that the actual voltage values are in the range $[0.98, 1.19]$, these error statistics strongly confirm that the W2V model can recover grid-wide voltage profiles accurately and consistently from weather inputs.

\subsection{Overall weather forecasting results}

\begin{table}[t]
\centering
\caption{RMSE comparisons between GAWF and GUWF}
\label{tab:GA_error_comparison}
\renewcommand{\arraystretch}{1.3}  
\setlength{\tabcolsep}{5pt}   
\resizebox{\columnwidth}{!}{%
\begin{tabular}{c|c|cccc}
\hline
\multirow{2}{*}{\shortstack[c]{Horizon\\(Lead Time)}} 
 & \multirow{2}{*}{Model} 
 & Temp. & Wind & GHI & Voltage \\
 &  & [°F] & [mph] & [W/m$^{2}$] & [$\times 10^{-3}$ p.u.] \\
\hline
\multirow{3}{*}{\shortstack[l]{~1-hr\\(1-hr)}} 
  & GUWF        & 1.561 & 1.086 & 29.947 & 4.079 \\
  & GAWF        & 1.586 & 1.084 & 29.950 & 3.859 \\
  & Error $\Delta$ [\%] & +1.602 & \textbf{-0.184} & +0.010 & \textbf{-5.393} \\
\hline
\multirow{3}{*}{\shortstack[l]{~3-hr\\(2-hr)}} 
  & GUWF        & 3.074 & 1.971 & 57.262 & 5.72 \\
  & GAWF        & 3.122 & 1.986 & 57.677 & 5.60 \\
  & Error $\Delta$ [\%] & +1.561 & +0.761 & +0.725 & \textbf{-2.150} \\
\hline
\multirow{3}{*}{\shortstack[l]{~6-hr\\(4-hr)}} 
  & GUWF        & 5.323 & 2.794 & 81.907 & 7.039 \\
  & GAWF        & 5.184 & 2.811 & 80.569 & 6.974 \\
  & Error $\Delta$ [\%] & \textbf{-2.611} & +0.608 & \textbf{-1.634} & \textbf{-0.923} \\
\hline
\multirow{3}{*}{\shortstack[l]{12-hr\\(7-hr)}} 
  & GUWF        & 6.870 & 3.508 & 80.756 & 8.050 \\
  & GAWF        & 6.961 & 3.525 & 82.840 & 7.969 \\
  & Error $\Delta$ [\%] & +1.325 & +0.485 & +2.581 & \textbf{-1.006} \\
\hline
\multirow{3}{*}{\shortstack[l]{~24-hr\\(14-hr)}} 
  & GUWF        & 9.349 & 4.048 & 97.961 & 8.776 \\
  & GAWF        & 9.393 & 4.043 & 98.997 & 8.599 \\
  & Error $\Delta$ [\%] & +0.471 & \textbf{-0.124} & +1.058 & \textbf{-2.017} \\
\hline
\end{tabular}%
}
\begin{flushleft}{\footnotesize Negative error values indicate improvements of GAWF compared to GUWF.}\end{flushleft}
\end{table}

Encouraged by the high accuracy attained by our W2V model, we next embed it as a fixed layer to obtain the GAWF and compare with the baseline weather forecaster (GUWF). Table~\ref{tab:GA_error_comparison} lists the RMSE values for the three weather features and the corresponding voltage predictions across five forecasting horizons. The error percentage values are also included to show the relative changes from GUWF to GAWF, with negative values indicating the improvements by GAWF highlighted by \textbf{bold} fonts. A clear trend in Table~\ref{tab:GA_error_comparison} is that GAWF reduces the voltage prediction error on all forecasting horizons. These voltage improvements are achieved with only marginal changes in weather forecast errors. This behavior suggests that the GAWF model is not optimized to merely improve the conventional weather forecasting accuracy. Rather, by incorporating the W2V predictor during training, GAWF learns weather forecasts that are better aligned with the downstream voltage prediction task. %it can support power system operations with more accurate modeling of voltage profiles.

The feature-wise RMSE trends further suggest that the downstream voltage gains are not equally sensitive to all weather features. Across most horizons, GAWF exhibits modest degradation in temperature and GHI, while the wind speed error remains comparable to or even slightly better than that of GUWF. Although the degradation of temperature forecasts is more pronounced than that of other features, it does not appear to negatively impact downstream voltage prediction, as GAWF still reduces voltage RMSE at every horizon. %{\ms{(This sentence seems too sudden)}} 
Notably, horizons with better wind forecasts tend to coincide with larger reductions in voltage RMSE. This pattern suggests that, among the three weather features, the wind speed is the most influential feature for voltage prediction in the studied system.

Across the horizons, the 6-hr horizon shows a distinct behavior with the smallest voltage improvement despite notable improvements in forecasting temperature and GHI. One possible explanation is the inherently lower predictability associated with this horizon. Specifically, a 4-hr lead time is known to cause low autocorrelation of GHI~\cite{BACHER20091772}, while a 6-hr forecasting horizon does not strongly match the daily cycle patterns in temperature and GHI compared to longer 12- or 24-hr horizons. Consistent with this interpretation, the 6-hr GHI error is already comparable to that of the 12-hr despite the shorter lead time and forecasting horizon. These inherent issues in 6-hr weather forecasting may have elevated baseline errors in temperature and GHI, leading to larger improvements in these features under GAWF.
%{\ms{(Not sure about causality of this sentence)}}

%further corroborates this interpretation: while temperature and GHI errors are notably reduced, this horizon yields the smallest improvement in voltage accuracy. This distinctive behavior

To more directly quantify the contribution of each weather feature to the voltage prediction improvement, we perform a \textbf{hybrid input analysis} using feature-wise substitution. We focus on the 1-hr and 24-hr horizons as representative short-term and day-ahead cases, with the best  accuracy at 1-hr versus significant uncertainty at 24-hr. For each weather feature, we replace the corresponding GUWF forecasts %{\ms{(feature?)}} 
with the GAWF counterpart while keeping the others unchanged, and assess the resulting voltage prediction. Table~\ref{tab:full_hybrid} summarizes the resulting hybrid voltage prediction errors and their percentage changes relative to the GUWF baseline from Table~\ref{tab:GA_error_comparison}. The results show that the improvement in wind speed forecasting is the dominant contributor to the reduction in voltage error at both horizons, especially in the 1-hr case. In contrast, the contributions from temperature and GHI are relatively small. Together with the earlier observations, we confirm that GAWF has strategically prioritized the forecasting of wind speed, the feature most influential for voltage prediction.

\begin{table}[t]
\renewcommand{\arraystretch}{1.2}
\caption{Comparison of Voltage RMSE via Hybrid Input Analysis}
%{\ms{Need to reduce the table size}}
\label{tab:full_hybrid}
\centering
\setlength{\tabcolsep}{5.0pt}
\begin{tabular}{c|c|c|c|c}
\hline
Horizon &
{\renewcommand{\arraystretch}{1.15}\begin{tabular}{@{}c@{}}
GUWF RMSE\\[-1pt]$[\times 10^{-3}\ \mathrm{p.u.}]$
\end{tabular}} &
{\renewcommand{\arraystretch}{1.15}\begin{tabular}{@{}c@{}}
Substituted\\[-1pt]GAWF Feature
\end{tabular}} &
{\renewcommand{\arraystretch}{1.15}\begin{tabular}{@{}c@{}}
Hybrid RMSE\\[-1pt]$[\times 10^{-3}\ \mathrm{p.u.}]$
\end{tabular}} &
{\renewcommand{\arraystretch}{1.15}\begin{tabular}{@{}c@{}}
Error $\Delta$\\[-1pt]{[\%]}
\end{tabular}} \\
\hline
\multirow{3}{*}{1-hr} & \multirow{3}{*}{4.079} & Temp. & 4.047 & -0.785 \\
\cline{3-5}
& & Wind & \textbf{3.926} & \textbf{-3.751} \\
\cline{3-5}
& & GHI & 4.038 & -1.005 \\
\hline
\multirow{3}{*}{24-hr} & \multirow{3}{*}{8.776} & Temp. & 8.735 & -0.467 \\
\cline{3-5}
& & Wind & \textbf{8.676} & \textbf{-1.139} \\
\cline{3-5}
& & GHI & 8.723 & -0.604 \\
\hline
\end{tabular}
%\begin{flushleft}{\footnotesize All RMSE values denote voltage RMSE.}\end{flushleft}
\end{table}

% \begin{flushleft}
% \footnotesize Error reduction rate is computed relative to baseline GUWF voltage RMSE values in Table~\ref{tab:GA_error_comparison}. \textcolor{red}{Include GUWF RMSE values as a separate column.}
% \end{flushleft}

\subsection{Performance comparisons across weather conditions}

\begin{figure*}[t]
    \centering
    \begin{tabular}[b]{@{}c@{}}
        \includegraphics[width=0.36\textwidth]{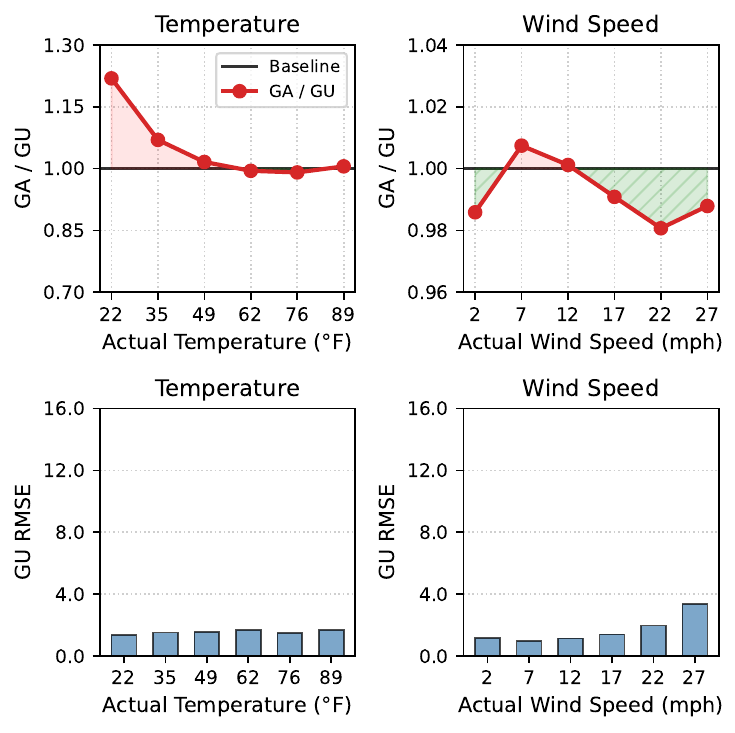} \\
        \small (a) 1-hr \\
    \end{tabular}
    \begin{tabular}[b]{@{}c@{}}
        \includegraphics[width=0.36\textwidth]{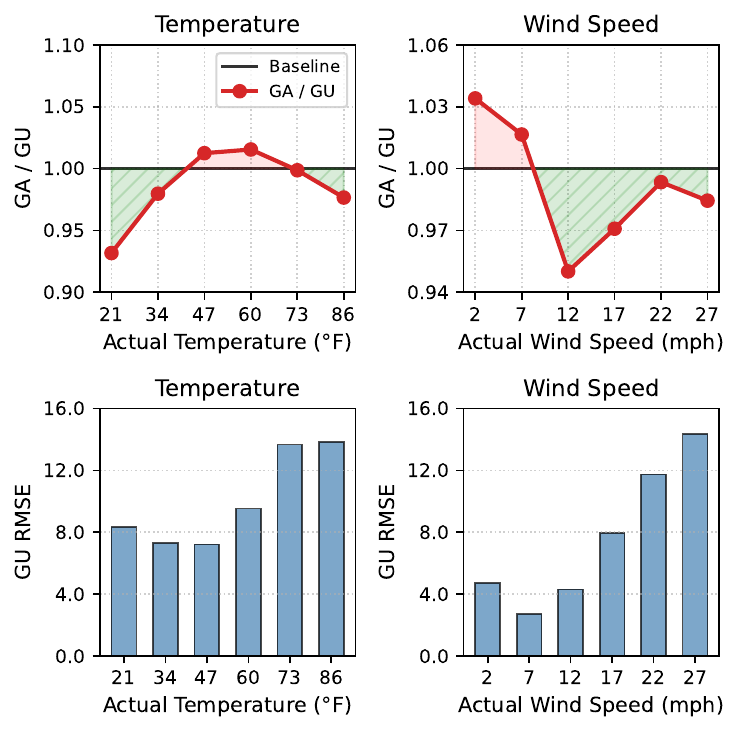} \\
        \small (b) 24-hr \\
    \end{tabular}
    \begin{tabular}[b]{@{}c@{}}
        \includegraphics[width=0.26\textwidth]{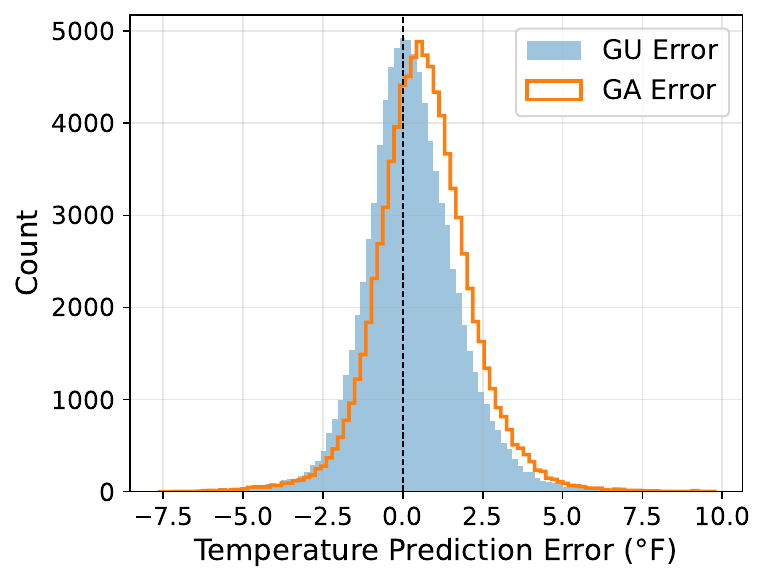} \\
        \vspace{0.9cm} \\
        \small (c) low-temperature forecasts at 1-hr\\
    \end{tabular}
    \caption{Weather forecast error variation across two horizons: (a) 1-hr and (b) 24-hr. Top row: RMSE ratio (GA/GU), where green shading indicates GA improvement (GA/GU $<$ 1), and red shading indicates GA degradation (GA/GU $>$ 1). Bottom row: baseline GU RMSE for each bin. Note that the y-axis scales for the GA/GU ratio differ between the two horizons. (c) Temperature forecast error distributions of GU and GA in the 1-hr low-temperature regime.}
    %{\hao reduce this fig, are all horizons discussed or needed? }
    \label{fig:error_distribution}
    \vspace{-0.3cm}
\end{figure*} 

To better demonstrate the changes by GAWF, we compare the weather forecasting performance in different ranges of weather conditions. We focus on the most representative 1-hr and 24-hr horizons, and GHI is omitted here due to the lack of clear weather-dependent patterns. Figs.~\ref{fig:error_distribution}(a) and (b) plot the RMSE ratio between GAWF and GUWF, along with the baseline GUWF RMSE for six bins of actual temperature and wind speed values.
At the 24-hr horizon, GAWF primarily improves the weather regions with large weather forecast errors, as shown in Fig.~\ref{fig:error_distribution}(b). In both temperature and wind speed, most bins wit GA/GU ratio below 1 coincide with those exhibiting higher GUWF errors. This pattern indicates that GAWF acts mainly as an error-correcting mechanism under long forecast horizons where weather uncertainty dominates.
A notable exception is the low-temperature regime, where GAWF provides improvement despite the relatively small GUWF errors. This behavior is meaningful since extremely low and high temperatures are often associated with stressed system loading conditions that can strongly affect grid voltages. Therefore, the results suggest that the proposed W2V-embedded GAWF does not merely follow the magnitude of weather forecasting error, but can also prioritize extreme weather conditions that are important for grid operations.

At the 1-hr horizon, GAWF is primarily driven by the objective of voltage prediction rather than reducing forecasting errors. As shown in Fig.~\ref{fig:error_distribution}(a), the baseline weather forecasts are already highly accurate across all bins, leaving limited room for direct error reduction. Accordingly, GAWF in this regime is better interpreted as being guided by the downstream voltage prediction objective. Consistent with this interpretation, GAWF degrades the forecasting performance in low-temperature bins while showing some improvement in high wind speeds.
To better understand this tradeoff, we further analyze the source of degradation observed in the low temperature, using the subset of samples from the first two bins in Fig.~\ref{fig:error_distribution}(a) (below $42^{\circ}\mathrm{F}$).
%{\hao why training samples? should it be test? } 
Using only this subset, we perform a hybrid input analysis by replacing only the GUWF temperature forecast by the GAWF forecast. This change reduces the voltage RMSE by \textbf{4.267\%} compared to the baseline using full GUWF outputs. Therefore, although GAWF produces larger temperature forecast errors in this subset, the modified temperature signal is more beneficial for the downstream voltage prediction. This result suggests the presence of a localized bias in the W2V model under low-temperature conditions. Because such samples account for only about 6\% of the training set, the model is likely less well calibrated in this regime. Fig.~\ref{fig:error_distribution}(c) further supports this interpretation by comparing the distributions of the GU and GA temperature forecast errors within the subset. The GAWF error distribution is shifted towards overprediction and a positive bias. This behavior indicates that GAWF intentionally introduces a positive temperature bias to compensate for the W2V bias and improve the final voltage prediction.  Note that this W2V bias is present at all horizons, but the 24-hr GAWF in  Fig.~\ref{fig:error_distribution}(b) has chosen to prioritize reducing the forecast error which is of higher concern than the bias issue. %This is because the latter incurs much worse forecast accuracy than the 1-hr one, which makes it a priority.  
 Collectively, these findings confirm that GAWF can adapt to the dominant error sources for downstream voltage prediction. Notably, GAWF could be sensitive to any W2V prediction error, underscoring the importance of W2V model development. Enhancing the accuracy and robustness of the W2V model using e.g., expanded training data, would further improve the practical applicability of this grid-aware framework.
%{\hao i thoguht we need a positive message on the importance of accurate W2V predictor?}

\subsection{Analysis of large voltage prediction error}

\begin{figure}[t]
    \centering
    \includegraphics[width=0.98\columnwidth]{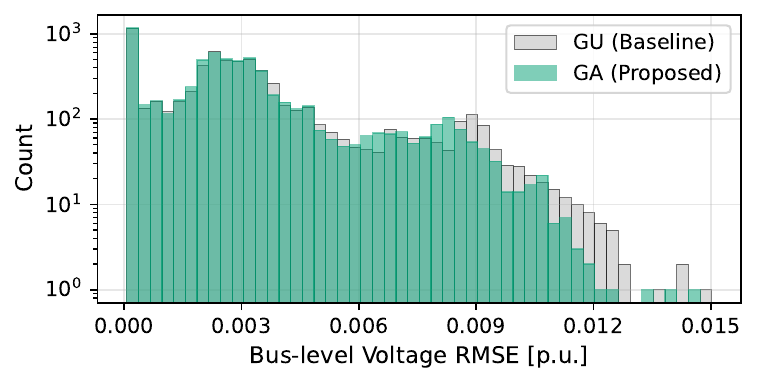}
    \vspace{-0.15cm}
    \caption{Histogram of bus-level voltage prediction RMSE at the 1-hr horizon under the baseline GUWF and the proposed GAWF.}
    \label{fig:large_vol_error}
\end{figure}

%{\ms{This paragraph also seems to deliver the same message with previous section. May need to shorten into one or two sentences and directly go into detailed analysis.}}
The primary benefit of GAWF is most clearly observed in reducing large voltage prediction errors. Fig.~\ref{fig:large_vol_error} compares the distributions of average voltage RMSE under GUWF and GAWF at the 1-hr horizon over the 6717 buses, showing that GAWF noticeably reduces the frequency of large voltage mismatches in the high-error tail.
%Since the 1-hr forecast is already very accurate for most samples, the two distributions largely overlap in the lower-error regime, with the main difference appearing in the high-error tail. GAWF noticeably reduces the frequency of large voltage mismatches, which confirms that the validity of the proposed GAWF for grid-level voltage impact analysis.
%To further investigate these samples where GAWF has the greatest impact, we analyze the large voltage error samples below.

To further analyze these large voltage errors, we select the top 5\% of all test samples ranked by the GUWF voltage RMSE. We focus on the 1-hr horizon so that the contribution of GAWF can be interpreted more clearly without the impact of long-term forecasting horizon.
%which allows us to identify the weather conditions linked to these deviations and to examine how GAWF adjusts its forecasts to alleviate them.  The 1-hr horizon is chosen for this analysis because GUWF is already highly accurate for most samples, so large voltage error events are more isolated and GAWF’s targeted adjustments are easier to discern than at longer horizons where baseline weather forecast errors are broadly elevated.
Table~\ref{tab:isolated_weather_error} compares the weather and voltage RMSE of GUWF and GAWF over this subset, along with baseline GUWF performance over the full test set. Notably, GAWF reduces voltage RMSE by 7.741\%, which is larger than the average improvement over the full test set (5.393\% in Table~\ref{tab:GA_error_comparison}). This improvement is primarily driven by enhanced wind forecasts, with only a marginal improvement in GHI and a degradation in temperature accuracy. The temperature degradation is consistent with the earlier low-temperature analysis: low-temperature samples constitute 21.0\% of this subset, compared with 15.8\% ratio in the full test set. This increased ratio makes the bias-compensation behavior of GAWF more prominent.

%{\hao how does it show comparison with the full test set?} 
These large error samples also corresponds to intrinsically more challenging meteorological conditions. Relative to the GUWF performance over the full test set, this subset exhibits 12.7\% higher wind forecast errors and approximately 1.5 times higher GHI forecast errors. In addition, 60.0\% of the samples in this subset occur during daytime hours (GHI $>$ 0), compared to only 43.0\% in the full test set. Because daytime GHI variability is inherently more difficult to capture than nighttime zero values, this higher concentration of daytime samples explains the large GHI errors in this subset. Despite these pronounced errors, GAWF achieves only marginal corrections in GHI, indicating that the voltage improvement cannot be primarily attributed to solar-related effects. %Therefore, we next examine the wind speed conditions in detail. 

%This improvement is primarily due to enhanced wind forecasts and a minor gain in GHI, although the temperature accuracy decreases. Several key characteristics of this subset help explain this result. First, the GUWF wind speed error in this subset is 12.7\% higher than that in the full test set listed in Table~\ref{tab:GA_error_comparison}, which likely has caused the large voltage mismatch. Second, low-temperature samples constitute 21.0\% of this subset, which is higher than the 15.8\% ratio in the full test set. This increased ratio together with the bias-compensation mechanism discussed earlier has led to the degradation in temperature forecast. Third, 60.0\% samples of this subset occur during daytime hours (GHI $>$ 0), compared to only 43.0\% in the full test set, where daytime GHI variability is inherently more difficult to capture than nighttime zero values. However, given the minor GHI improvement under GAWF, its impact on the voltage prediction is likely limited.
%\textcolor{blue}{I think higher error values are observed in GHI than wind. What's the impact of GHI?}

\begin{table}[!t]
\renewcommand{\arraystretch}{1.2}
\caption{Weather and Voltage RMSE Comparison for Isolated Large Voltage Error Samples (1-hr)}
\label{tab:isolated_weather_error}
\centering
\setlength{\tabcolsep}{5.0pt}
\begin{tabular}{c|c|c|c|c}
\hline
Comparison & Temp. & Wind & GHI & Voltage \\ 
Type & [°F] & [mph] & [W/m$^{2}$] & [$\times 10^{-3}$ p.u.] \\ 
\hline
GU RMSE (Full Set) & 1.561 & 1.086 & 29.947 & 4.079 \\
\hline
GU RMSE (Large Error) & 1.628 & 1.224 & 45.357 & 10.515 \\
GA RMSE (Large Error) & 1.667 & 1.211 & 45.256 & 9.701 \\
\hline
Error $\Delta$ [\%] & +2.396 & \textbf{-1.062} & \textbf{-0.223} & \textbf{-7.741} \\
\hline
\end{tabular}
\end{table}

\begin{figure}[t]
  \centering
  \includegraphics[width=\columnwidth]{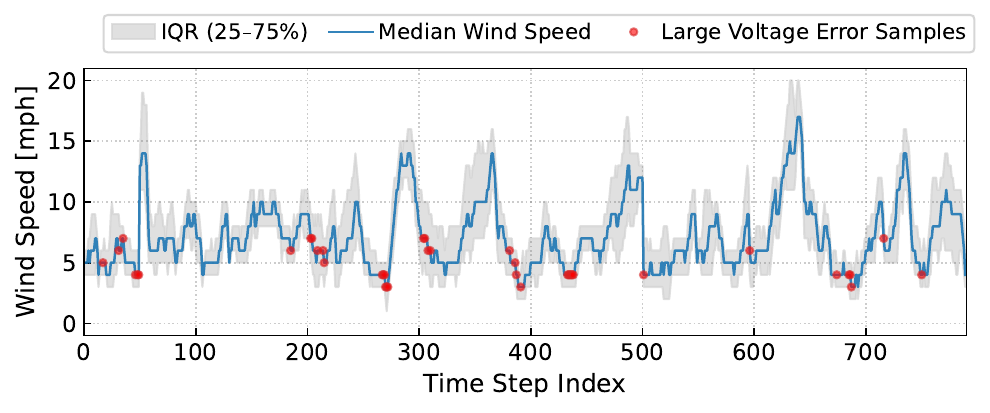}
  \caption{Wind speed distribution across 701 weather grid points for all test time steps. Red markers indicate time steps with large voltage errors selected by RMSE.}
  \label{fig:isolated_wind}
\end{figure}

Instead, the dominant contributor to improved voltage prediction remains wind. Fig.~\ref{fig:isolated_wind} shows the system-wide wind speed conditions for the full test set using median and interquartile range (IQR) across all 701 weather grid points, with the large voltage error samples indicated in red markers.
%To ensure robustness to evaluation criteria, we isolate large voltage error samples using three complementary metrics: RMSE, MAE, and entry-wise maximum ($\ell_\infty$ norm) error, which emphasize different aspects of the error distribution.
%\textcolor{red}{(Do wee need to incorporate that we have compared three metrics?) - No, we don't need to. The other metrics are removed.}
%RMSE is sensitive to large deviations, MAE reflects the average magnitude of errors, and the entry-wise maximum captures peak errors.
%Despite these differences, the isolated samples identified by the three metrics substantially overlap; Fig.~\ref{fig:isolated_wind} shows the RMSE-selected subset. This consistency indicates that large voltage deviations occur under specific wind speed conditions, rather than being artifacts of the choice of metric. 
Interestingly, the large voltage errors tend to coincide with low-wind conditions, typically near the valleys of the median wind speed profile immediately before the ramp-up periods. Such transition periods are characterized by rapid and volatile wind fluctuations, which are well known to be difficult to forecast accurately---a challenge widely recognized in the context of forecasting wind ramping events \cite{11068153}. Under these conditions, GUWF incurs relatively large wind forecast errors, while GAWF provides targeted corrections that substantially reduce the resulting voltage mismatch, as previously discussed in Table~\ref{tab:isolated_weather_error}. This voltage-centric view also complements the weather-regime analysis in the previous subsection. On average, the wind-related benefit of GAWF at the 1-hr horizon appears more visible in the higher wind bins. However, the large-voltage-error analysis reveals that the most operationally important gains are concentrated in transition events, especially in low-wind periods immediately preceding ramp-ups. Because such events are relatively sparse, their significance is partly obscured when errors are averaged over coarse weather bins.

%{\ms{it'd be better to focus only on this subsection, the different or new conclusion compared to the previous subsection}}
In summary, our numerical studies validate that the proposed GAWF approach consistently improves voltage prediction accuracy across the range of forecasting horizons and weather conditions. GAWF effectively prioritizes the forecast of wind speed, the meteorological feature that is the most volatile yet strongly coupled with bus voltage prediction. Furthermore, GAWF exhibits adaptive behavior according to the underlying forecast uncertainty, transitioning from error-focused correction at longer horizons to voltage-driven adjustment at shorter horizons. More importantly, analysis of large voltage error samples reveals that GAWF concentrates its improvements during critical weather transitions, particularly during system-wide low-wind conditions preceding ramp-ups. These findings demonstrate that the proposed grid-aware approach produces weather forecasts that are inherently more beneficial for downstream grid operations.
% \textcolor{blue}{why low wind speed is consequential for grid stability? I think high wind power and extreme ramps are the concerns of grid stability.}

% This interpretation is further corroborated by the hybrid input analysis applied to the isolated large voltage error subset. As shown in Table \ref{tab:isolated_hybrid}, wind speed achieves the largest improvement in voltage prediction accuracy. A small improvement is also observed when temperature forecasts are replaced, likely reflecting residual bias-correction effects in the low temperature samples. 

\section{Conclusion and Future Directions}
\label{sec:conclusion}

We have developed a direct weather-to-voltage (W2V) predictive modeling framework and demonstrated its effectiveness in achieving grid-aware weather forecast at the weather–grid nexus. We first design the W2V surrogate model that can predict bus voltages directly from weather features and efficiently provides gradients for end-to-end learning. Its compact autoencoder architecture with dual decoders retains weather information while ensuring accurate and generalizable voltage predictions, and PCA-based encoder initialization further enhances training stability.
%{\hao this part didn't highlight the design improvements.} 
Building on this, the W2V predictor is applied as a regularizer, yielding a grid-aware weather forecaster (GAWF). Numerical experiments show that the GAWF consistently produces weather forecasts that are more beneficial to downstream voltage prediction than the conventional GUWF while accepting a slight degradation in overall weather accuracy. Instead of uniform meteorological precision, the GAWF prioritizes the weather features and conditions critical to grid states, most notably by enhancing wind speed accuracy during sudden, system-wide wind drops preceding ramp-ups to mitigate high voltage errors. This demonstrates that our grid-aware approach effectively learns the weather–grid coupling to better serve grid operational needs.

%{\hao long, need to reduce}
Exciting future research directions include designing the neural W2V architecture using graph filters to be consistent with the geospatial mapping between weather and grid locations. Furthermore, the W2V predictor can support new applications in identifying grid-critical weather conditions or perturbations that can facilitate more proactive and risk-aware grid operations in the face of extreme weather events. 

%\textcolor{blue}{Check refs [2], [10], [13], [17]}
%\ifCLASSOPTIONcaptionsoff
%\fi
\bibliographystyle{IEEEtran}
\bibliography{mybib}
\end{document}